\documentclass[aps,floats,amssymb,twocolumn,prb]{revtex4-2}
\usepackage{calc}
\usepackage{psfrag}
\usepackage{graphicx}
\usepackage{color}
\usepackage{bm}

\begin{document}

\title{PH Pfaffian intra and inter correlations in the quantum Hall bilayer}

\author{M. V. Milovanovi\'c$^{1}$ and S. Djurdjevi\'c$^{2,3}$ } 
\affiliation{$^{1}$ Scientific Computing Laboratory, Center for the Study of Complex Systems, Institute of Physics Belgrade, University of Belgrade,
Pregrevica 118, 11080 Belgrade, Serbia}
\affiliation{$^2$  University of Montenegro, Faculty of Natural Sciences and Mathematics, D\v zord\v za Va\v singtona bb, 81000 Podgorica, Montenegro}
\affiliation{$^3$ Faculty of Physics, University of Belgrade, Studentski trg 12, 11158 Belgrade, Serbia}

\begin{abstract}
PH Pfaffian topological phase may exist in a uniform system due to strong Landau level (LL) mixing according to theoretical predictions based on the Son - Dirac composite fermion theory. Numerical investigations in the presence of large  LL mixing are limited due to numerical complexities, when taking into account at least one more LL. Because of this, we apply the same field theoretical approach to the quantum Hall bilayer at total filling factor equal to one, for which many numerical studies exist. The most advanced in [\onlinecite{zfs}] predicts an intermediate phase (for intermediate distances between layers) with an even-odd effect. According to our approach, the intermediate phase represents a mixed negative-flux $p$-wave pairing  i.e. coexisting intra (PH Pfaffian in each layer) and inter (a la PH Pfaffian) pairing correlations. This again underlines a necessity for strong entanglement with additional degrees of freedom, i.e. at least one more (additional) LL in the search for a stable PH Pfaffian phase in a single layer. Based on the analogy with the bilayer physics we propose a PH Pfaffian wave function that resides in two LLs.
\end{abstract}

\maketitle


The FQHE (fractional quantum Hall effect) [\onlinecite{tsg}] phenomena are usually explained by taking a single LL projection - a projection of the Hilbert space of the problem to a single LL. This is not always justified with respect to experiments (the so-called LL mixing may be large), but captures efficiently the special commensuration of the number of electrons with respect to the number of flux quanta through the system, that leads to the stability of states and quantization of the Hall conductance.

The FQHE at 5/2 [\onlinecite{rw}] is an example when LL mixing may be decisive in selecting a state with a particular order;  the most prominent candidate state from numerical experiments is anti-Pfaffian [\onlinecite{ap1, ap22, rez}] (a particle-hole conjugate of Pfaffian state [\onlinecite{moor}]), but the experiments [\onlinecite{w1,w2}]  point out to the presence of the so-called PH Pfaffian topological order. The PH Pfaffian state is proposed as the state that will reflect the PH symmetry  (the symmetry under exchange between particles and holes) of an isolated half-filled LL, but, in a way paradoxically, it is expected to be stabilized by disorder and LL mixing [\onlinecite{zf}] . That this is the case may be recovered by considering the Dirac composite fermion theory [\onlinecite{son}] - a theory that incorporates the PH symmetry of an isolated half-filled LL and examining possible pairing channels [\onlinecite{avm}]. The PH Pfaffian order is present when the Dirac mass - the PH symmetry breaking agent is considerable. 

The bilayer problem, in a first approximation, is a problem with two LLs, usually the same one - the lowest LL (LLL) in each layer,  that are degenerate. The two degenerate LLs may differ in general as discussed in  [\onlinecite{haf}]. The two degenerate LLs we may view as an extreme case of LL mixing, which is completely justified as in the case of the graphene bilayer [\onlinecite{pa}], or may simulate (in an approximate way) the effect of LL mixing, if we consider one of the layers at higher chemical potential i.e. the cyclotron energy. 

In this work we reexamine the quantum Hall bilayer physics at filling factor one i.e. when each layer is half-filled. By assuming the projection to the LLL in each layer, we apply the Dirac composite fermion theory and examine pairing instabilities. They are the same as in the single layer, but may be classified as possibilities for inter and intra pairings.  The most relevant to the existing numerical work are those that describe the bilayer at intermediate distances. We identify them as negative flux $p$-wave (a la PH Pfaffian) inter and PH Pfaffian (intra) pairing correlations that are associated with the presence of mass in each layer due to the interaction between layers, and thus in the absence of the PH symmetry inside layers. 

In the treatment of the quantum Hall bilayer we will apply the usual Chern-Simons (CS) approach, but also consider its version with Dirac composite fermions (CFs) - fields, which takes into account the PH symmetry (the symmetry under exchange of particles and holes), if it is present, like in the case of an isolated half-filled LL. The CS approach via gauge field(s) incorporates the Coulomb interaction among underlying electrons by connecting electron with its correlation hole. We may in a way speak about an excitonic instability (although correlation hole is not an independent degree of freedom) and, like in the Laughlin case at $\nu = 1/3$, bosonic exciton (exciton = electron + correlation hole) condensation. At $\nu = 1/2$ correlation hole is of bosonic nature, and the resulting exciton is fermionic, and  a way to include the Coulomb repulsion is to set each two electrons (more precisely excitons) apart by considering a $p$-wave pairing.  These $p$-wave correlations are present in the non-relativistic CS descriprtion as shown in [\onlinecite{gww}]  via the ``statistical interaction" - influence of gauge field on particles, and the source of  the gauge field are the same particles.  This picture is much cleaner and complete if we explicitly include the PH symmetry of an isolated half-filled LL, and thus consider without bias, on equal footing, electrons and holes, though, on one hand, this may result in non-existant phenomena because we  are (it seems) artificially doubling the degrees of freedom, but, on the other hand, the ``excitonic" physics, Coulomb repulsion may be captured more efficiently. In Fig. 1 presented are results for $p$-wave instabilities, based on the Dirac CS field theory for the half-filled LL, in the presence of a mass term, which is a PH symmetry breaking parameter.  The internal gauge field, i.e. its change $ \delta {\bf a}$ from the uniform mean-field value, is described by
\begin{equation}
	\frac{1}{2} \frac{\vec{\nabla} \times {\bf a}}{2 \pi} = c^{\dagger} c + d^{\dagger} d,
\end{equation}
i.e. sources are what we may call composite fermion (CF) - composite of electron and correlation hole, represented by field $c$ (a component of the Dirac spinor) and composite hole - a composite of hole and its ``correlation particle", represented by a field $d$ (another component of the Dirac spinor) [\onlinecite{dg}]. It is interesting to note that with the two degrees of freedom we have possibility for ``intra" pairing (Pfaffian and anti-Pfaffian) and  ``inter" pairing (PH Pfaffian). 
\begin{figure}[th]
\centerline{\includegraphics[width=9.7cm]{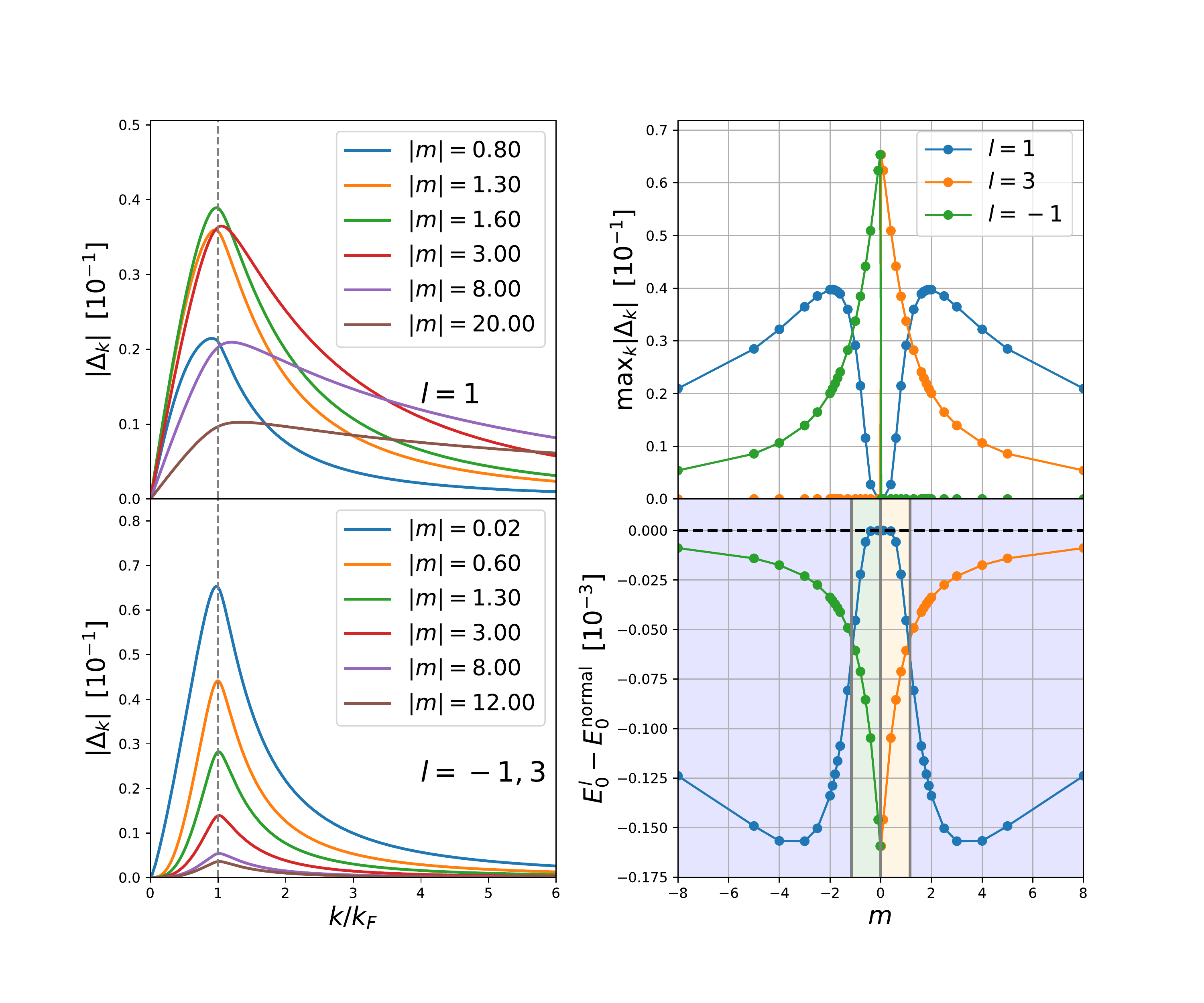}}
\vspace*{8pt}
\caption{The solution of the self-consistent BCS problem. Left column: radial direction $k$-dependent pairing amplitude for various values of $m$.
Channel $l = 1$ solution (PH Pfaffian) only depends on $|m|$, while $l = 3$ (anti-Pfaffian) and $l = -1$ (Pfaffian) channel solutions are symmetric with the sign-flip of $m$.
Upper right panel: dependence of the maximum of the pairing amplitude on $m$ (always found at the Fermi level $k_F$).
Lower right panel: total energy of the different pairing solutions compared to the normal state energy.
Gray vertical lines denote the transition between different channels. Color in the background corresponds to the energetiically favorable channel at the given $m$ - a measure of Landau level mixing. The color of lines: Pfaffian - green, anti-Pfaffian - orange, PH  Pfaffian - blue. From the Ref. [\onlinecite{avm}].}
\end{figure}

The quantum Hall bilayer at total filling factor one is an old subject [\onlinecite{e1, e2}]; when layers are close to each other i.e. $d$ - the distance between layers, $ d \lesssim l_{B}$ (magnetic length), due to the Coulomb repulsion, a real excitonic instability occurs between electrons and holes that are in the opposite layer. This is captured by the following ground state wave function,
\begin{equation}
\Psi _{(111)} =  \prod_{i<j} (z_{i \uparrow} - z_{j \uparrow})  \prod_{k<l} (z_{k \downarrow} - z_{l \downarrow}) \prod_{m<n} (z_{m \uparrow} - z_{n \downarrow}),
\end{equation}
i.e. the ``(1,1,1)" state. If we confine the physics to the lowest LLs (LLLs)  of both layers, we expect for large distances the physics of separate, two half-filled LLLs and the Dirac CF description to be valid in each layer.
When layers are close, the ground state is very much the (1,1,1) state, and we can describe this system by a (non-relativistic) CS theory [\onlinecite{lf}]: 
\begin{eqnarray}
{\cal L} & =  &  \sum_{\sigma} \Psi_{\sigma}^{\dagger} ( i \partial_t - A_0 - a_0^\sigma ) \Psi_{\sigma} \nonumber \\
&& - \sum_{\sigma} \frac{1}{2 m}  \Psi_{\sigma}^{\dagger} ({\bf p} - {\bf A} - {\bf a}_{\sigma} )^2 \Psi_{\sigma}  - \frac{1}{2 \pi}  a_{\uparrow} \partial a_{\downarrow} \nonumber \\
&& + interactions
\end{eqnarray}
By varying $a_0^{\sigma}$, $\frac{\delta {\cal L}}{\delta a_0^\sigma} =0$,  we get,
\begin{equation}
- \Psi_{\sigma}^{\dagger} \Psi_{\sigma} - \vec{\nabla} \times {\bf a}_{-\sigma} = 0.
\end{equation}
Thus the effective magnetic field is
\begin{eqnarray}
\frac{\vec{\nabla} \times {\bf A}^{\sigma}_{\rm eff}}{2 \pi}& =& \frac{\vec{\nabla} \times {\bf A}}{2 \pi} + \frac{\vec{\nabla} \times {\bf a}^{\sigma}}{2 \pi} \nonumber \\
&=& \rho - \rho_{-\sigma} = \rho_{\sigma}.
\end{eqnarray}
Thus CS fermions are at the integer filling factor equal to one (each), in each layer. We introduce a shift in variable $a$, $ {\tilde a}_{\sigma} = a_\sigma + A$,
\begin{eqnarray}
{\cal L} & =  &  \sum_{\sigma} \Psi_{\sigma}^{\dagger} ( i \partial_t  - {\tilde a}_0^\sigma ) \Psi_{\sigma} \nonumber \\
&& - \sum_{\sigma} \frac{1}{2 m}  \Psi_{\sigma}^{\dagger} ({\bf p}  - {\tilde {\bf a}}_{\sigma} )^2 \Psi_{\sigma}  - \frac{1}{2 \pi}  {\tilde a}_{\uparrow} \partial {\tilde a}_{\downarrow} \nonumber \\
&& +  \frac{1}{2 \pi}  ({\tilde a}_{\uparrow} + {\tilde a}_{\downarrow}) \partial A - \frac{1}{2 \pi}  A \partial A .
\end{eqnarray}
(We omitted interactions in the last expression.) 

To connect this description with the one at large distances (in which an effective Dirac physics is expected), let's consider  a Dirac system Lagrangian, $ {\cal L}_D $ that in the limit of large masses becomes effectively ${\cal L}$. Thus we expect and assume that effective description at $d$ large continously connects with the one at  $d$ small.  
\begin{eqnarray}
{\cal L}_D & =  &  \sum_{\sigma}     i {\bar \chi}_\sigma \gamma^\mu (\partial_\mu + i {\tilde a}_\mu^\sigma ) \chi_\sigma  - m \sum_{\sigma} {\bar \chi}_\sigma \chi_\sigma           \nonumber \\
&&  - \frac{1}{2 \pi}  {\tilde a}_{\uparrow} \partial {\tilde a}_{\downarrow} \nonumber \\
&& +  \frac{1}{2 \pi}  ({\tilde a}_{\uparrow} + {\tilde a}_{\downarrow}) \partial A - \frac{1}{2 \pi}  A \partial A .
\end{eqnarray}
Let's consider the case with $m = 0$ and apply the duality transformation. Here, for a fixed $\sigma$,  ${\tilde {\bf a}^{\sigma}}$ plays the role of a uniform background field and thus we apply this transformation  assuming small fluctuations  in  ${\tilde {\bf a}^{\sigma}}$ i.e.
$\vec{\nabla} \times {\tilde {\bf a}}_{\sigma} \approx \frac{B}{2}$ where $ B = \frac{\vec{\nabla} \times {\bf A}}{2 \pi} $. Also these are not  neutral Dirac systems in background fields, but each system is at the filling factor one  i.e. half-filled  in the $n = 1$ LL (not $n = 0$ LL). Thus if the usual duality transformation brings out half-filled LL physics of an isolated, usually LLL, in this case $n = 1$, and we need to add CS terms to get the right Hall conductance of the subsystems, which are not neutral.    Thus we are introducing new spinors, $\eta_\uparrow$ and $\eta_\downarrow$, and  in the new representation,
\begin{eqnarray}
{\cal L}_d (m=0) & =  &  \sum_{\sigma}     i {\bar \eta}_\sigma \gamma^\mu (\partial_\mu + i  b_\mu^\sigma ) \eta_\sigma            \nonumber \\
&& +  \sum_\sigma \frac{1}{2 \pi}  b_{\sigma} \partial {\tilde a}_{\sigma} - \frac{1}{4 \pi} \sum_\sigma   {\tilde a}_{\sigma} \partial {\tilde a}_{\sigma}\\
&&  - \frac{1}{2 \pi}  {\tilde a}_{\uparrow} \partial {\tilde a}_{\downarrow} \nonumber \\
&& +  \frac{1}{2 \pi}  ({\tilde a}_{\uparrow} + {\tilde a}_{\downarrow}) \partial A - \frac{1}{2 \pi}  A \partial A .
\end{eqnarray}
As the response due to the variation of external field we have
\begin{equation}
j_e^\mu \equiv - \frac{\delta {\cal L}_d}{\delta A_\mu} = \frac{\partial A}{\pi} - \frac{\partial {\tilde a}_{\uparrow} + \partial {\tilde a}_{\downarrow}}{2 \pi}.
\end{equation}
Once again we see that for $ j_e^0 \equiv \rho_e = B$, in the mean field, and applying the symmetry under exchange of $\uparrow$ and $\downarrow$ we get 
\begin{equation}
\frac{\vec{\nabla} \times {\tilde {\bf a}}^{\uparrow}}{2 \pi} = \frac{\vec{\nabla} \times {\tilde {\bf a}}^{\downarrow}}{2 \pi} = \frac{B}{2}.
\end{equation}
Also, from the equations of motions,  $\frac{\delta {\cal L}_d}{\delta b_\mu}  = 0$, we have
\begin{equation}
- j_{\sigma}^\eta + \frac{\partial {\tilde a}^{\sigma}}{2 \pi} = 0,
\end{equation}
and from $\frac{\delta {\cal L}_d}{\delta {\tilde a}_\sigma}  = 0$,
\begin{equation}
\frac{\partial b^{\sigma}}{2 \pi} - \frac{\partial {\tilde a}^{\sigma}}{2 \pi} - \frac{\partial {\tilde a}^{-\sigma}}{2 \pi} + \frac{\partial A^{\sigma}}{2 \pi} =0.
\end{equation}
Thus in the mean field $ \frac{\vec{\nabla} \times {\bf b}^{\sigma}}{2 \pi} = 0$. It is not hard to see that if we integrate out field $ a_\uparrow - a_\downarrow $ we get $ b_\uparrow = b_\downarrow = b $. Integrating out $ a_\uparrow + a_\downarrow $ gives the following effective action:
\begin{eqnarray}
{\cal L}(m=0) & =  &  \sum_{\sigma}     i {\bar \eta}_\sigma \gamma^\mu (\partial_\mu + i b_\mu ) \eta_\sigma             \nonumber \\
&& +  \frac{1}{4 \pi}  (b + A) \partial (b + A) - \frac{1}{2 \pi}  A \partial A .
\end{eqnarray}
Thus, even in the case when $m = 0$, which we can identify with the one when the layers are infinitely apart, we do not have two independent systems, but effective correlations between two layers. We extrapolate this description to the case when $m \neq 0$ by the following Lagrangian density,
\begin{eqnarray}
{\cal L} & =  &  \sum_{\sigma}     i {\bar \eta}_\sigma \gamma^\mu (\partial_\mu + i b_\mu ) \eta_\sigma             \nonumber \\
&& + m \sum_\sigma {\bar \eta}_\sigma \eta_\sigma \nonumber \\
&& +  \frac{1}{4 \pi}  b \partial b + \frac{1}{2 \pi}  b \partial A - \frac{1}{4 \pi}  A \partial A ,
\end{eqnarray}
where $m$ here may be a function of the old $m$, certainly monotonically increasing function of the old $m$. The mass term comes with + sign in order to eliminate the $ \frac{1}{4 \pi}  b \partial b $ term in the Pauli-Villars regularization so that the Hall conductance of the whole system is $ 1 \cdot \frac{e^2}{h} $, given by the last term in the Lagrangian.
We are in the Zhang's representation (picture) of quasiparticles because
\begin{equation}
j_e^\mu \equiv - \frac{\delta {\cal L}}{\delta A_\mu} = \frac{\partial A}{2 \pi} + \frac{\partial b}{2 \pi},
\end{equation}
and, from the equations of motions,  $\frac{\delta {\cal L}_d}{\delta b_\mu}  = 0$, we have
\begin{equation}
- j_\uparrow^\eta  - j_\downarrow^\eta + \frac{\partial b}{2 \pi} +   \frac{\partial A}{2 \pi}= 0,
\label{eqden}
\end{equation}
i.e. the variation of  $ j_e = j_e^\uparrow + j_e^\downarrow $ and $  j_\uparrow^\eta  + j_\downarrow^\eta $ with respect to $ \frac{\partial A}{2 \pi}$ is up to a sign,  $ \frac{\partial b}{2 \pi}$, i.e. the same.

We can solve for ${\bf b}$ using equation (\ref{eqden}) in terms of the density of the $\eta$ quasiparticles and investigate the Cooper channels of the statistical interaction,
\begin{equation}
V_{\rm st} = - \sum_\sigma {\bar \eta}_\sigma {\vec \gamma} {\bf b} \eta_\sigma ,
\end{equation}
that may lead to various Pfaffian intra-paired or inter ($p$-wave, $\cdots$ ) paired states.

Also, the limit of small and intermediate mass we identify or consider as cases when layers are away from each other (not close), because $m$ represents a parameter which tell us how much the particle-hole symmetry inside each layer is spoiled from the case of isolated layers when $m = 0$.

Following the same steps as in Ref. [\onlinecite{avm}]  we may arrive at the conclusion that for the intermediate values of $m$, the relevant Cooper channel interaction i.e. dominant (possibility for) pairing is of the following form
\begin{eqnarray}
V_{Cch} & = & \sum_{{\bf k}, {\bf p}, \sigma } V_{{\bf k}, {\bf p}} a_{{\bf k}, \sigma}^\dagger  a_{-{\bf k}, \sigma}^\dagger  a_{-{\bf p}, \sigma}  a_{{\bf p}, \sigma} \nonumber \\
&&  \sum_{{\bf k}, {\bf p}} 2  V_{{\bf k}, {\bf p}} a_{{\bf k}, \uparrow}^\dagger  a_{-{\bf k}, \downarrow}^\dagger  a_{-{\bf p}, \downarrow}  a_{{\bf p}, \uparrow} ,
\end{eqnarray}
with
\begin{equation}
 V_{{\bf k}, {\bf p}} = \frac{2 \pi}{8 V} \frac{1}{E_p E_k} \{ - 4 m |{\bf k}| |{\bf p}| \frac{ i \sin\{\theta_p - \theta_k \}}{|{\bf k} - {\bf p}|^2} \},
\end{equation}
which describes PH Pfaffian (intra) correlations, inside each layer, and, with the same kind of vorticity, intercorrelations between the layers. 

If we assume the intra pairing, the mean field Hamiltonian is of the following form,
\begin{eqnarray}
K_{\rm eff}^a  & = & \sum_{{\bf k}, \sigma} \xi_{{\bf k}} a_{{\bf k}, \sigma}^\dagger  a_{{\bf k}, \sigma} \nonumber \\
&&  + \sum_{{\bf k}, \sigma } \frac{1}{2} ( \Delta_k^a  a_{{\bf k}, \sigma}^\dagger  a_{-{\bf k}, \sigma}^\dagger      + (\Delta_k^a)^*  a_{-{\bf k}, \sigma}  a_{{\bf k}, \sigma} ),
\end{eqnarray}
and in the case of inter pairing we may write,
\begin{eqnarray}
K_{\rm eff}^e  & = & \sum_{{\bf k}, \sigma} \xi_{{\bf k}} a_{{\bf k}, \sigma}^\dagger  a_{{\bf k}, \sigma} \nonumber \\
&& + \sum_{{\bf k}, \sigma }  ( \Delta_k^e  a_{{\bf k}, \uparrow}^\dagger  a_{-{\bf k}, \downarrow}^\dagger      + (\Delta_k^e)^*  a_{-{\bf k}, \downarrow}  a_{{\bf k}, \uparrow} ).
\end{eqnarray}
In the mean field treatment we find $\Delta_k^e = \Delta_k^a $, and that the ground state energies are the same i.e. inter and intra pairing are equally likely, irrespective of the value of mass (for intermediate values of distance). The solution is given in Fig. 1 in blue color. This independence of mass i.e. distance between the layers of the ratio between inter and intra pairing is certainly consequence of our previous approximations. As the distance is increasing we can expect that intra pairing is becoming more likely (intra Coulomb repulsion is becoming more pronounced) and the share of intra pairing is increasing in a possible mixed state in which inter and intra pairing coexist. 

This mixed pairing ground state description is consistent with the numerical results of Ref.  [\onlinecite{zfs}]. The odd-even effect that is observed in the data for intermediate distances, is consistent with the stability of states with even number of particles in each layer with respect to states with odd number number of particles  in each layer;  if we have an even number in each layer we have to break a pair to make a transfer to another layer. Also the persistance of the superfluid stiffness (density) (in this study of finite systems) in this mixed state may be connected with the necessary persistance of inter pairing correlations and pairs which represent the inter-layer ordering at intermediate distances. This was already observed in [\onlinecite{msr}], in the work that proposed and detected for the first time inter $p$-wave pairing. (This persistance of the superfluid stiffness should be a consequence of algebraic ODLRO [\onlinecite{mdp}] in this regime which is marked by the absence of the Goldstone mode. This finite stiffness is a finite size effect. But this regime, for a finite (not zero) interval of distance, with persistant interlayer correlations, without LRO, cannot be a finite size effect as also
evidenced in recent experimental data [\onlinecite{ei}]. A question may be whether intra PH Pfaffian correlations may survive the thermodynamic limit, and, also, the presence of disorder in experiments.) 

The presented CS description of the quantum Hall bilayer, in conjunction with the numerical results in [\onlinecite{zfs}] gives further evidence that the PH Pfaffian (intra) correlations (in a layer) require the presence of correlations or entanglement with the second layer or a second (another) LL. The mixed state phase should be bounded at a small distance, $d_{c1}$, by a critical state in which all Cooper pairs are pairs between CFs in different layers, and at a large distance, $d_{c2}$, by two PH Pfaffian states, each state describing (intra) pairing in its corresponding layer. (This direct product state should be in the same phase of two composite fermion Fermi-liquid-like state [\onlinecite{bal, mish, avm, ym}] and $d_{c2}$ may be at infinity.) Although the two-component Dirac CS formalism introduces seemingly artificial degrees of freedom, its prediction of the inter and intra opposite vorticity - PH Pfaffian correlations and instabilities in the bilayer system is fully consistent with the available numerical data. This gives support for relevance and motivates further investigation of PH Pfaffian correlations in a single layer in the presence of strong LL mixing. 

The doubling of degrees of freedom in the Son's Dirac composite fermion theory introduces a possibility for PH Pfaffian, but if $m =0$, and we are describing isolated half-filled LL, the possibility is artificial - it is based and includes an additional (artificial) degree of freedom i.e. it describes the excitonic binding of electrons and holes (which are not independent degrees of freedom). Remarkably , this is reflected in the field theory which has no PH Pfaffian instability at $m =0$ as we can see in Fig. 1.  We need at least one additional LL (``layer") to induce PH Pfaffian correlations, in order to justify physically the excitonic pairing, which in the one-component, low-energy limit (of the Dirac theory near Fermi level) becomes the $p$-wave pairing of CFs. This is similar to the two-component, quantum Hall bilayer physics, but certainly two LLs do not represent two equivalent subsystems of the bilayer. We may expect finite density of electrons in the higher LL, in an ideal, strong-coupling PH Pfaffian construction in which all CFs are paired, and subsystems are not identical like in the construction proposed in [\onlinecite{dff}]. (This strong-coupling, ideal form for Pfaffian can be defined in an isolated LL (Moore-Read construction [\onlinecite{moor}] ).)

Based on the quantum Hall bilayer analogy we propose a PH Pfaffian wave function, that, in the long-distance limit - with no projection applied, is of the following form:
\begin{equation}
\Psi_{\rm PHPf} =  \prod_{i<j} (z_i - z_j )^2 Pf \{ \frac{({\bar z}_i + {\bar z}_j )}{({\bar z}_i - {\bar z}_j )} \},
\label{phh}
\end{equation} 
where $Pf$ denotes the antisymmetrized product (collection) of Cooper pairs described by $\frac{({\bar z}_i + {\bar z}_j )}{({\bar z}_i - {\bar z}_j )}$ , of spinless - indistiguishable electrons.
A description of  details of the analogy  and an analysis of the proposed wave function can be found in the Appendix.
By adding $({\bar z}_i + {\bar z}_j )$ factor to each Cooper pair $(i,j)$, first of all we are transformig neutral sector (not charged sector) and we  are at, overall, the same filling factor 1/2.  In fact, by associating this factor to each pair, we are putting (on average) one electron of the pair in the higher, second LL. For a fixed configuration of pairs, there are factors of the form $\prod_{ \{i = 1\} }^{\{N/2\}} {\bar z}_{ \{i\} }$ (where curly brackets denote those $N/2$ electrons  that are chosen from $N$ of them - a fixed partition of $N$), and that is telling us that half of the electrons are in the second LL. Thus the filling factor of the LLL is 1/4, and it is the same with the second LL. (The total filling factor is 1/2.)

The topology of the state is still of $p$-wave superconductor (up to charge modes). We can see that by considering edge excitations and bulk quasielectron excitations that should generate them. In the case of Pfaffian [\onlinecite{mr}], to extract Majorana mode from the bulk quasihole excitations the following Pfaffian identity was crucial:
\begin{equation}
Pf(a_i - a_j ) = 0
\end{equation}
if $a_i $, $ i = 1, \ldots, P, P > 2 $ and even, to eliminate spurious states in the Pfaffian case. In that case $\{a_i  \}$ were the coordinates of electrons, $\{z_i \}$. In the case of the proposed PH Pfaffian, due to factors $({\bar z}_i + {\bar z}_j )$  and complex conjugated coordinates in the construction of bulk quasielectrons, for detailes see [\onlinecite{mj}], the complex  $\{a_i  \}$ are $\{{\bar z}_i^2  \}$, and we can use the Pfaffian identity to again extract Majorana edge states (but of opposite vorticity).



Thus, when LL1 (the first excited level) and LL2 are degenerate the PH Pfaffian construction,  (\ref{phh}), may be the ground state. As we raise the chemical potential of LL2, the construction may evolve into states of the following form,
\begin{equation}
  \prod_{i<j} (z_i - z_j )^2 A\{ Pf \{ \frac{({\bar z}_i + {\bar z}_j )}{({\bar z}_i - {\bar z}_j )} \}   \prod_{\{p\}} \exp\{i {\vec k}_p {\vec r}_p \} \},
\label{ph}
\end{equation}
(where the product of plane waves describes electrons that do not pair)
i.e. composite fermions may transfer into the lower LL by forming the Fermi sea (we omitted the projection of the Fermi sea - composite fermion liquid (CFL)  part).  (We are moving away from the ideal case but the topology should stay the same [\onlinecite{rg}].) The final result of the raising of the chemical potential of LL2, would be a CFL state that will compete with the Pfaffian, anti-Pfaffian superposition in the case of the isolated half-filled LL1. 

Thus, we may ask ourselves whether a construction that we described may exist in the case of two LLs. And whether  it is relevant for the explanation of the FQHE at filling factor 5/2. Numerical explorations may give the answers,  a similar study as the one in Ref. [\onlinecite{pa}]. There LL0 and LL1 were considered degenerate (in the context of the graphene bilayer), and, in fact, from this study we can conclude, that in the case of the (``nonrelativistic") single layer, even if we neglect the cyclotron energy, the (spin polarized) electrons (or holes) concentrate (group) in one LL at filling factors 1/2 and 3/2. So dominant is the exchange effect, so stable is the correlated state, of Fermi liquid and Pfaffian kind, that they exist even in the case of degenerate LLs (when particles ``have more space").  (We may say the system is polarized - electrons tend to occupy a particular LL.) 

In the case of FQHE at 5/2 we may (similarly) assume that the spin quantum Hall effect inside LL0 is so stable and strong (electrons are ``polarized" in a special way) that we may concentrate our attention to the two relevant LLs : LL1 and LL2. Also see the reference [\onlinecite{dm}] for an argumentation for considering only two LLs in the case of the PH Pfaffian. If the system supports PH Pfaffian topological phase, electrons should not group in a single LL in the ground state (because the PH Pfaffian cannot exist in a single LL, the most recent numerical studies are in Refs. [\onlinecite{rph}] and [\onlinecite{ym}]).

In Fig. 2 are Coulomb matrix elements - Haldane pseudopotentials in the problem with two LLs: LL1 and LL2. 
\begin{figure}[th]
\centerline{\includegraphics[width=8cm]{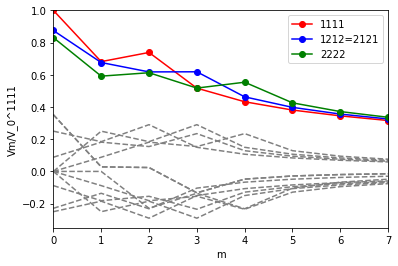}}
\vspace*{8pt}
\caption{Matrix elements of generalized Haldane pseudopotentials for the LL1 and LL2, for Coulomb interaction relative to $V_0^{1111}$.}
\end{figure}
In the case of the reduced Hamiltonian with only elements: $V^{1111}, V^{2222}$, and $V^{1212} = V^{2121}$, the system is still polarized in LL1 due to the strong exchange effect [\onlinecite{zpc}]. But if we include  $V^{1221} = V^{2112}$ in Ref. [\onlinecite{haf}] was demonstrated that the ground state is unpolarized Haldane-Rezayi state. Also this reference states that the depolarization of this kind occurs only in the LL1, LL2 degenerate case and not in the LL0, LL1 case. The question is what  is the ground state when we include  all matrix elements in the LL1, LL2 case. If we stabilize PH Pfaffian correlations, then we need to test whether they persist if we increase the chemical potential of LL2. The complete diagonalization that we plan for future work may also indicate whether PH Pfaffian correlations support an incompressible state.

While we were finishing the preparation of this manuscript, another related work, [\onlinecite{sh}], appeared on arXiv,  that explored the physics of bilayer at filling factor one. Our conclusions overlap in the way that PH Pfaffian correlations can be viewed as a $s - wave$ pairing of composite fermion and composite hole - the concept that was introduced in Ref. [\onlinecite{dg}] which gives a microscopic derivation of the Dirac composite fermion theory. The geometry of the numerical work in  [\onlinecite{sh}] is the geometry of sphere, which may be biased towards particular instabilities, while the geometry of the numerical experiment in [\onlinecite{zfs}] is the torus geometry (which does not have that bias but may be disadvantageous in other ways). The predictions of both references overlap in the most important and interesting - intermediate region for which we showed that an effective field-theoretical description is also possible.

We would like to thank A. Balram, D. Feldman, N. Regnault, and Z. Papic for correspodence and discussions.

\appendix
\section{The bilayer and single layer wave functions}

Due to the inclusion of mass(es) $m$ in the description of each layer in the case of bilayer, Dirac composite fermions are becoming ordinary (nonrelativistic) composite fermions of the usual CS description, for considerable $m$ in the intermediate region. Thus, for the wave function that will desribe the inter negative-flux  pairing, in the long distance, we expect the following form,
\begin{eqnarray}
\Psi_{e}(z_{i_1 \uparrow}, \ldots , z_{i_{N/2} \uparrow}, z_{j_1 \downarrow}, \ldots , z_{j_{N/2} \downarrow})= \nonumber \\
= Det \{ \frac{1}{{\bar z}_{i \uparrow} - {\bar z}_{j \downarrow}}\}\cdot \nonumber \\
 \cdot\prod_{i_1 < i_2 }^{N/2} (z_{i_1 \uparrow} - z_{i_2 \uparrow})^2 \prod_{j_1 < j_2 }^{N/2} (z_{j_1 \uparrow} - z_{j_2 \uparrow})^2  ,
\end{eqnarray}
where $Det $ denotes an antisymmetrized collection of Cooper pairs described by $\frac{1}{{\bar z}_{i \uparrow} - {\bar z}_{j \downarrow}}$. As in the field-theoretical description we differentiate between $\uparrow$ and $\downarrow$ electrons, $ N_{\uparrow} = N_{\downarrow} = N/2$, where $N$ is the total number of electrons. We can make the construction totallly antisymmetric i.e. electrons indistingushable in the layer index by considering,
\begin{eqnarray}
\sum_{\langle i_1 , \ldots , i_{N/2} \rangle} \Psi_{e}(z_{i_1 \uparrow}, \ldots , z_{i_{N/2} \uparrow}, z_{j_1 \downarrow}, \ldots , z_{j_{N/2} \downarrow})\cdot\nonumber\\
\cdot |\uparrow_{i_1}, \ldots , \uparrow_{i_{N/2}}, \downarrow_{j_1}, \ldots , \downarrow_{j_{N/2}}\rangle ,
\end{eqnarray}
i.e. summing over all partitions of electrons into the two distinctive groups of $\uparrow$ and $\downarrow$ electrons. We can see that the orbital and pseudospin part are entangled; the expression is not a direct product of the orbital and pseudospin part. 

In a direct analogy with the bilayer case, when all electrons participate in inter pairing, we may consider a wave function for equally populated two LLs, at total filling $1/2$, in which compsite fermions pair in the same way as in the bilayer case,
\begin{equation}
\sum_{\langle i_1 , \ldots , i_{N/2} \rangle} Det \{ \frac{1}{{\bar z}_{i} - {\bar z}_{j}}\}  \prod_{k < l }^{N/2} (z_{k} - z_{l})^2  {\bar z}_{i_1 }  \cdots {\bar z}_{i_{N/2}}
\label{phdet}
\end{equation}
The wave function corresponds to the case of the  LLL $ \equiv$  LL0 and the first excited LL, LL1, and the whole expression requires a projection to those two LLs. In this case the Jastrow-Laughlin correlations do not distinguish layer index, and we can rewrite the wave function in the following way,
\begin{equation}
\Psi_{\rm PHPf} =  \prod_{i<j} (z_i - z_j )^2 Pf \{ \frac{({\bar z}_i + {\bar z}_j )}{({\bar z}_i - {\bar z}_j )} \},
\label{aphh}
\end{equation}
the same as (\ref{phh}) in the main text. Here $Pf$ denotes the antisymmetrized product (collection) of Cooper pairs described by $\frac{({\bar z}_i + {\bar z}_j )}{({\bar z}_i - {\bar z}_j )}$ , of spinless - indistiguishable electrons, and we use
\begin{equation}
Pf\{ \frac{1}{({\bar z}_i - {\bar z}_j )} \} \sim \sum_{\langle i_1 , \ldots , i_{N/2} \rangle} Det \{ \frac{1}{{\bar z}_{i} - {\bar z}_{j}}\} .
\end{equation}
The factor $({\bar z}_i + {\bar z}_j )$ in (\ref{aphh}) corresponds to a triplet Cooper pairing in the pseudospin (the LL index) language. In general, the factor ${\bar z}$ places - modifies and projects the single particle wave function $z^{m} \exp\{- \frac{1}{4} |z|^2 \}$ of the LLL ($l_B$ (magnetic length) =1), $ m = 0,1, \ldots, N_\phi -1 $, $N_\phi $ is the number of flux quanta through the system, into the LL1 wave function $z^{m} {\bar z}  \exp\{- \frac{1}{4} |z|^2 \}$. Thus we can interpret  the presence of factor  $({\bar z}_i + {\bar z}_j )$ in the Cooper pair description as an expression of the fact that the center of mass coordinate of the Cooper pair is in the LL1 (higher LL) with the center of mass angular momentum equal to zero. Thus, though $ z ( {\bar z}) $ coordinate is affected by translation, the correlations built in the wave functions are translationally invariant, and ${\bar z}$ acts as a pseudospin degree of freedom. This is more transparent in (\ref{phdet}).

If the construction were of the following form,
\begin{equation}
 \prod_{i<j} (z_i - z_j )^2 Pf \{ \frac{1}{({\bar z}_i - {\bar z}_j )} \} \sum_{\langle i_1 , \ldots , i_{N/2} \rangle} {\bar z}_{i_1 }  \cdots {\bar z}_{i_{N/2}},
\label{phfl}
\end{equation}
(i.e.  a direct product of orbital and pseudospin part) the projection to the two LLs would be effectively projection to the LLL of the usual PH Pfaffian construction, that would lead to a compressible state with correlations of a composite fermion liquid [\onlinecite{mish, ym}]. The projection of (\ref{aphh}) may lead to an incompressible state.

\end{document}